\documentclass[a4paper,twocolumn]{esapub2005} 

\usepackage{times}
\usepackage{natbib}
\usepackage{graphicx}

\begin{document}

\pagestyle{empty}

\title{IMAGING THE DUST TRAIL AND NECKLINE OF 67P/CHURYUMOV-GERASIMENKO} 

\author{Jessica Agarwal}
\affil{MPI-K, Saupfercheckweg 1, 69117 Heidelberg (Germany), Email: jessica.agarwal@mpi-hd.mpg.de}
\author{Hermann B\"ohnhardt}
\affil{MPS, Max-Planck-Str. 2, 37191
  Katlenburg-Lindau (Germany), Email: boehnhardt@mps.mpg.de}
\author{Eberhard Gr\"un$^{1,}$}
\affil{HIGP, 1680 East West Road POST 512c, Honolulu, HI 96822 (USA), Email: eberhard.gruen@mpi-hd.mpg.de}

\newcommand{\btx}{\textsc{Bib}\TeX}

\maketitle

\begin{abstract}
We report on the results of nearly 10 hours of integration of the dust trail and
neckline of comet 67P/Churyumov-Gerasimenko (67P henceforth) using the Wide Field Imager at the
ESO/MPG 2.2m telescope in La Silla. The data was obtained in April 2004 when
the comet was at a heliocentric distance of 4.7 AU outbound. 67P is
the target of the {\em Rosetta} spacecraft of the European Space Agency.
Studying the trail and neckline can contribute to the quantification of \mbox{mm-sized} dust grains released by the comet. We describe
the data reduction and derive lower limits for the surface
brightness. 
In the processed image, the angular separation of trail and neckline is
resolved. We do not detect a coma of small, recently emitted grains.
\end{abstract}
\section{INTRODUCTION}
The trajectories of cometary dust particles are -- to first order --
determined by their emission speed relative to the nucleus and by the ratio
$\beta$ of solar radiation pressure to solar gravity. 
Both quantities decrease with increasing particle size when the latter is
large compared to the wavelengths of sunlight. 
\newline
Large (mm/cm-sized) dust grains remain close to the orbit of their
parent comet for many revolutions around the sun, appearing to the observer as a long, line-shaped structure, the comet's {\em dust trail}. 
The emission of such particles is thought to be the principal mechanism
by which a comet loses refractory mass to the interplanetary dust environment
\cite{sykes-walker1992a}. Trails of eight short-period comets were first
observed with IRAS in 1983 \cite{sykes-lebofsky1986,sykes-hunten1986}, one of
them being that of 67P.
\newline
Of similar shape is the {\em neckline}
\cite{kimura-liu1977} which consists of dust
released from the comet at a true anomaly of 180$^\circ$ before observation.
In our case this corresponds to emission in mid-July 2002, roughly a month before perihelion passage. 
An observer close to the comet's orbital plane will see the
neckline as a thin bright line, slightly inclined  
with respect to the projected orbit. 
Particles in the neckline are younger and on average
smaller than in the trail.
\newline
Comet trails and necklines are best studied when separated from smaller dust
grains. The latter are released at higher relative speeds and are subject to
stronger radiation pressure. They disperse in space on timescales of
weeks to months after their emission, and their presence is not expected in the
vicinity of an inactive comet far from the sun. 
In this circumstance lies the appeal of observing a cometary trail or neckline 
at large heliocentric distance, even despite the then
fainter surface brightness. 
\newline
In Section~\ref{sec:data}, we describe and discuss the processing of the
raw data. This is followed in Section~\ref{sec:interpretation} by the
interpretation of the obtained image paying special attention to the
discrimination between dust trail and neckline. The results are summarised in
Section~\ref{sec:summary}.
\section{DATA ACQUISITION AND PROCESSING}
\label{sec:data}
\begin{figure*}[t]
\includegraphics[clip,width=\textwidth]{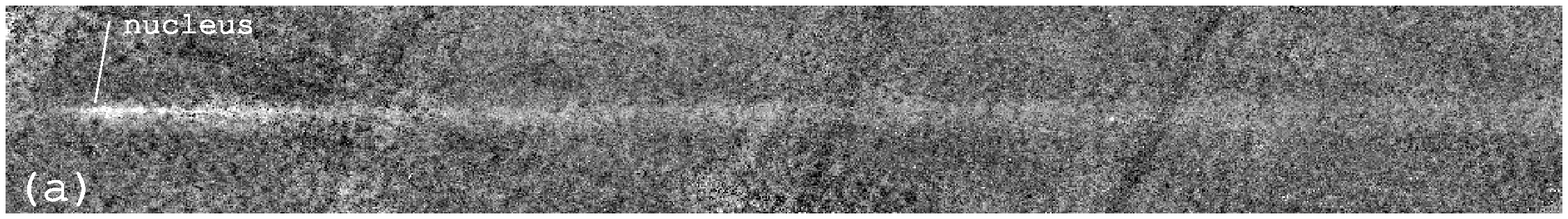}
\includegraphics[clip,angle=270,width=\textwidth]{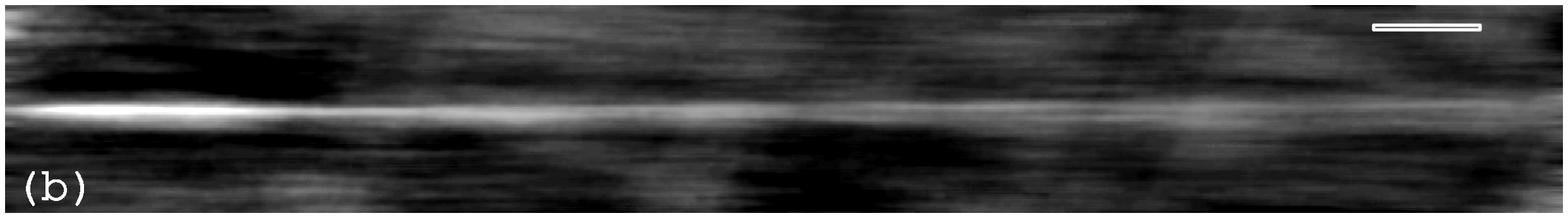}
\caption{(a) Unfiltered image rotated by 26.5$^\circ$ clockwise. 
  (b) Same image, each pixel being replaced by the average over a
  neighbourhood of 200 pixels (140$^{\prime\prime}$)
  parallel and 10 pixels (7$^{\prime\prime}$) perpendicular to the trail
  axis after removal of the nucleus. The filtering window is
  indicated in the upper right corner. 
  The size of the images is 35$^{\prime}$\,x\,4.7$^{\prime}$ each. For orientation see Fig.~\ref{fig:synsyn}.
  The inclined stripes are remnants of the overscan regions between
  individual CCDs. 
}
\label{fig:images}
\end{figure*}

67P was observed in April 2004 with
the Wide Field Imager (WFI) at the ESO/MPG 2.2m telescope in La Silla. The
heliocentric and geocentric distances of the comet were \mbox{4.7 AU} and \mbox{3.7 AU}, respectively. 
The total integration time was \mbox{9.8 h}. 45 minutes were
done on 2 April; the remaining time was split equally over the four
consecutive nights of 18 -- 21 April. We discarded the exposures taken on 18 April,
which were highly contaminated by stray light from a star of 4th
magnitude outside but close to the instrument field of view (FOV).
The remaining data comprises 50 images of \mbox{540 s} exposure
time each. The physical width of one pixel is 15 $\mu$m corresponding to 0.238$^{\prime\prime}$.  
In order to maximise sensitivity, 3$\times$3 on-chip pixel binning
was used. Each image is a mosaic of 8 CCDs covering a total FOV of
\mbox{34$^{\prime}$$\times$33$^{\prime}$}. No filter was applied.
\newline
The data processing was done with IRAF. A thorough discussion of the
peculiarities of WFI data and the individual steps of their reduction is given
in \cite{erben-schirmer2005}. The raw images were bias-subtracted, flatfielded,
corrected for airmass, and a mean sky-brightness was subtracted. They were then
average-combined while masking stellar objects as described by \cite{ishiguro-sarugaku_subm}. In order to increase the
signal-to-noise ratio (SNR), a spatial averaging filter was applied to the
resulting image. 
An approximate flux calibration was achieved using field stars.
\newline
The raw images were characterised by strong fringing, an
interference artefact arising in blue-optimised thin-layered CCDs when
observing in red wavelengths \cite{erben-schirmer2005}. In our data, the fringing pattern was spatially
constant, and its amplitude to first order proportional to the background
intensity level. 
Neither twilight nor dome flatfield exposures were used, because the
images were taken without filter and the spectral properties of the night sky
are different from those of the twilight sky or lamp. Instead, superflats were
built from the science data directly. This was possible because due to
jittering, stellar objects were in different positions on the CCD in
successive exposures. The superflats were obtained by median averaging over several normalised
images. Thus bright objects were excluded from the combined image while
instrument-specific features remained. An optimally smooth and fringe-free
background in the flatfielded images was achieved if using five
consecutive exposures per superflat.
%
To make different images comparable a mean sky level was subtracted from each. 
Airmass correction was done assuming a mean extinction coefficient for La
Silla of 0.15 mag/airmass. 
\newline
In the following, we call the data thus obtained the ``corrected single
images''. They were subsequently processed in two different manners 
in order to, first, allow for an approximate flux calibration using field
stars and, second, obtain the final image of the trail.
For the flux calibration, SNR was increased by averaging over all corrected
single images of a given night with such offsets that stars would superpose.  
In the combined images, aperture photometry of a set of ``solar-type'' field
stars in the FOVs of the images was done. Stars were considered as
``solar-type'' when their B-R and R-I filter colours in the USNO-B1.0
catalogue were compatible with solar values within the accuracy of the
catalogue of $0.25$ mag \citep{monet-levine2003}. Only stars with
R-magnitudes fainter than $17.7$ could be used, because brighter ones were
saturated in the raw data. 
As R-magnitude we used the mean of the two values given in the catalogue.
For all stars $j$ fulfilling the above criteria, the integral flux $I_j$ in
the combined image was measured (in arbitrary
units). By plotting the catalogue R-magnitude $M_{\rm cat} (j)$ versus $-2.5 \log_{10} (I_j)$ and fitting a linear
relation 
\begin{equation}
M_{\rm cat} (j) = M_0 - 2.5 \log_{10} (I_j)
\end{equation}
to the data points, the calibration offset $M_0$ was deduced. This procedure
was exercised for all four nights independently. The derived values of $M_0$ were consistent within the regression errors. We
used their mean for further calculations: $M_0 = 32.34 \pm 0.02$. 
We obtain surface brightness values ranging from $27.0$ to $28.4$
mag/arcsec$^2$ (in R) for the trail and an average nucleus magnitude of $21.7 \pm 0.1$.
\newline
The final trail image was obtained by averaging over all corrected single
images with such offsets as to align the comet. Due to the
relative motion of comet and background objects, the latter would appear in the
combined image as short, dashed lines often considerably brighter than the trail.
In order to exclude such objects from
being considered by the averaging procedure, object masks were applied. The object mask for a given night was created
with the IRAF routine {\tt objmasks} for the star-wise combined images used already
for the aperture photometry. Saturated or otherwise bad
pixels and the spaces between CCDs were masked as well.
\newline
The resulting combined
image has a reasonably smooth background and the trail is easily visible (Fig.~\ref{fig:images}(a)). However, the low mean
SNR of 0.6 per individual pixel precludes a quantitative analysis of this image. SNR can be improved by
applying a spatial averaging filter which replaces each pixel by the average
of its \mbox{$m\times n$} neighbours (IRAF routine {\tt boxcar}). The price to
pay is loss of spatial resolution. 
The size of the averaging window should be smaller than the
characteristic dimension of the object. Since the trail is very extended
in the direction parallel to its axis while narrow perpendicular to it,
we chose a rectangular window much larger in the parallel
direction than in the perpendicular one. 
To properly apply the filter, the image was first rotated 
aligning the trail to the x-axis (rotation by
$26.5^\circ$ clockwise). In addition, the nucleus was removed, i.e. replaced by an interpolation over the
surrounding pixels. We used a filtering window of 200 pixels (140$^{\prime\prime}$) parallel
and 10 pixels (7$^{\prime\prime}$) perpendicular to the trail axis, increasing SNR per pixel by a factor of 45.
The resulting image is shown in Fig.~\ref{fig:images}(b). 
\newline
Before embarking upon a quantitative analysis of the data, a certain problem
arising from the flatfielding process must be addressed. 
The difficulty originates from the fact
that the surface brightness of the trail was much less than the statistical
variation of the background. Moreover, the jittering pattern had not been
optimised in direction and amplitude to ensure that trail information could be
completely excluded from the superflats. 
\newline
The superflats were
constructed by median averaging over five consecutively taken exposures. In
each of these, a given object is found in a different
position. Putting it the other way round, if we consider a certain pixel in five
consecutive exposures, it will contain a given object at maximum once. This will be discarded by the median filter, and hence the resulting
image is free of bright objects. 
The method fails if
the object is less bright than the statistical fluctuation of the
background. Since in this case the object-containing pixel will not be significantly
brighter than the other four, there is a non-vanishing probability that the median
of the five pixels will happen to
be the one bearing the object. 
In the limit of a
very faint object, the chances of having it in the median pixel
approach 20\% which is the probability for one out of five identically
distributed pixels to assume the median value. This means that up to 20\% of
the pixels in the concerned regions of the superflat must be expected to bear
trail information. 
This information is lost from the original image on division by the superflat,
and the resulting surface
brightness will be 20\% too low on average. 
An accurate estimate of the loss is difficult,
and any attempt at a quantitative correction would be highly speculative. It remains
always true that the measured brightness is a lower limit. 
\newline
To nevertheless enable a quantitative comparison with simulated
images, we propose to include the details of the flatfielding process into the
simulation.
The simulated image will then contain (to first order)
the same artefacts as the WFI image and should hence be comparable to it.
\section{INTERPRETATION}
\label{sec:interpretation}
\begin{figure*}[t]
\includegraphics[clip,width=\textwidth]{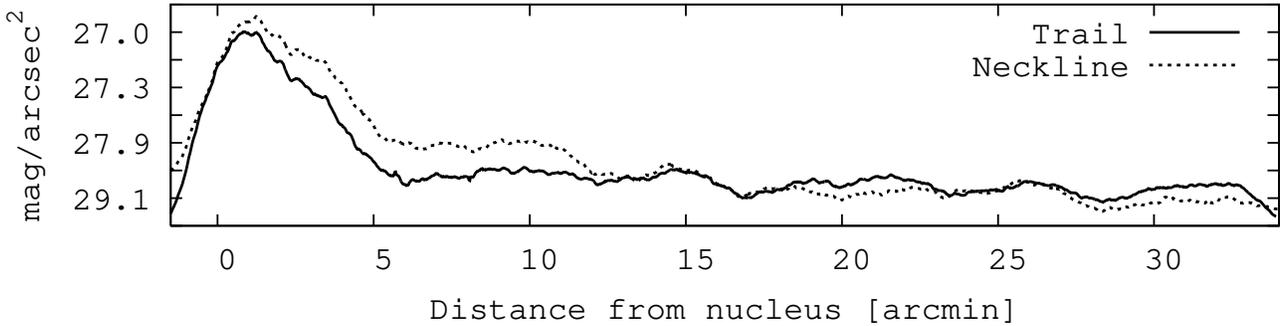}
\caption{Lower limit brightness profiles along trail and neckline. The scale
is linear in intensity but labelled with the corresponding logarithmic
R-magnitudes. The profiles are not independent of each other
because trail and neckline overlap, especially near the nucleus (see Fig.~\ref{fig:images}(b)). The apparent
shift of the peak to the right of the nucleus is a result of the spatial
averaging filter, the brightness dropping more steeply to the left of the
nucleus than to its right.}
\label{fig:profiles}
\end{figure*}
Taking a closer look at the filtered image
Fig.~\ref{fig:images}(b), a splitting of the line-shaped structure can be discerned. The
gap between the two parts widens with increasing distance from the
nucleus. 
We have ascertained that the
splitting does not result from combining images taken in different nights:
It remained if only data acquired in a single night was
used, and the predicted position angle of the trail did not change
significantly over the period of observation. 
Hence we presume that the splitting is real. 
\newline
Using the model described
in \cite{agarwal-mueller2006_inpress}, we find that the expected position
angles of trail and neckline are $296.9^\circ$ and $296.1^\circ$, respectively
(measured counterclockwise from north).
In Fig.~\ref{fig:images}, the difference in position angle between the two
branches is $0.8^\circ \pm 0.2^\circ$ which complies well with the
anticipated separation of trail and neckline. The same is true for the mean
position angle of the feature which is $296.5^\circ$. Hence we interpret the
upper branch in Fig.~\ref{fig:images} as the dust trail and the lower one as
the neckline. 
Trail and neckline are visible up to the edge of the FOV. The length of the
orbit section covered is $35^{\prime}$, corresponding to $1.1^\circ$ in mean anomaly.
\newline 
The different ages of dust in the trail and neckline
imply that at a given distance from the nucleus trail particles are
expected to be larger on average than material in the neckline. If we assume
that the size distribution of dust leaving the comet did not
change with time (apart from its dependency on the strength of gas drag), the
projectional separation of trail and neckline provides us with two
manifestations of the same quantity. This puts an additional constraint to any
model attempting to reproduce the data.
\newline
Fig.~\ref{fig:profiles} shows intensity profiles along the axes of trail
and neckline. They are characterised by a pronounced peak around the
nucleus and a rather uniform brightness distribution at distances beyond $5^{\prime}$
from the nucleus. 
According to simulation, the surface brightness in the neckline decreases
significantly with growing nucleus distance while it is rather uniform along
the trail.
Since the influence of radiation pressure 
decreases with particle size, larger grains remain
closer to the nucleus. Therefore, the bright peak in Fig.~\ref{fig:profiles} is likely due to mm/cm-sized
particles emitted around perihelion in 2002. 
\begin{figure}[h]
\label{fig:synsyn}
\includegraphics[clip,width=\columnwidth]{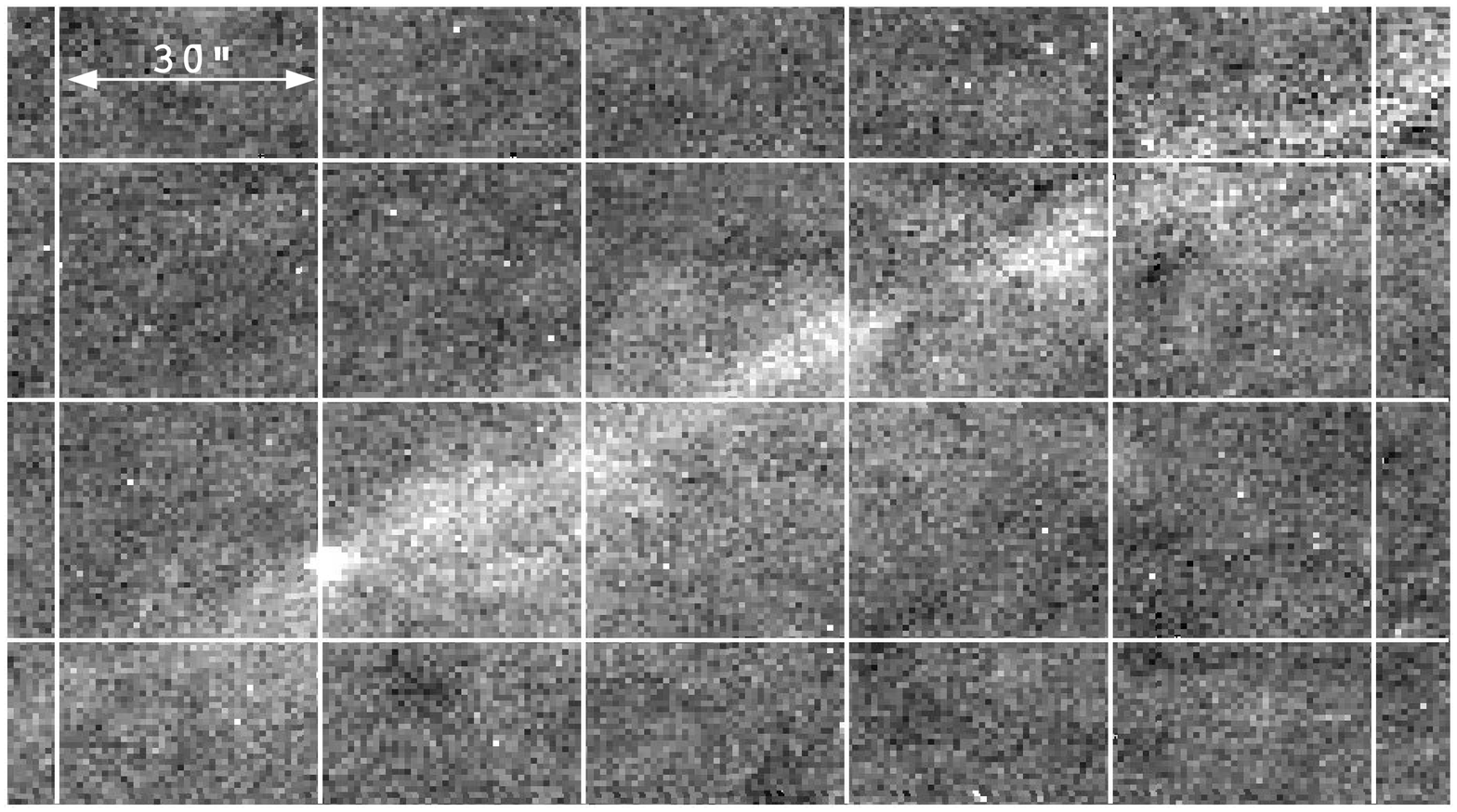}
\includegraphics[clip,width=\columnwidth]{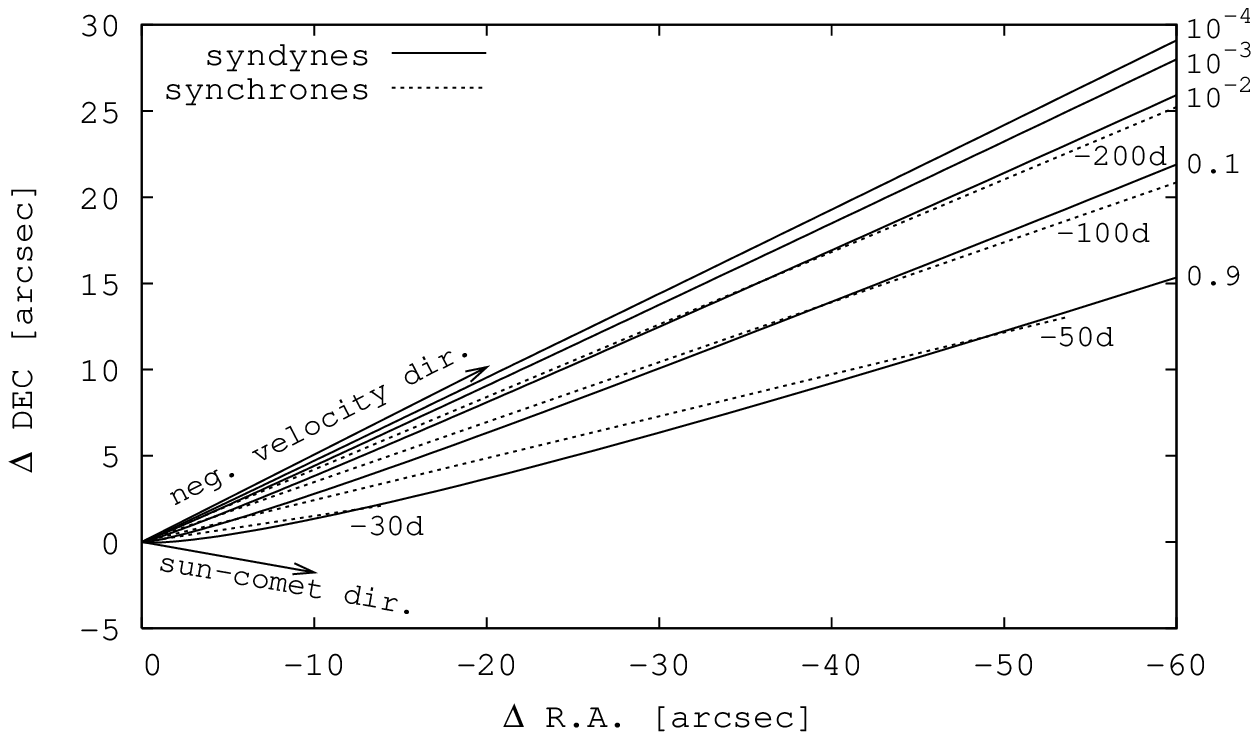}
\caption{Top: Enlarged view of the nucleus and adjacent region in
  Fig.~\ref{fig:images}(a) before rotation. Bottom: A plot of synchrones and
  syndynes for 19 April 2004. Synchrones are the positions of particles
  of different $\beta$ emitted at fixed times, labelled in the plot in days
  before observation. Syndynes show the positions of particles of given
  $\beta$ (indicated at the right margin) released at varying dates. The
  negative orbital velocity vector of the comet corresponds to its
  projected orbit and trail axis. Syndynes of large and synchrones of old particles in the
  image converge towards this direction. The extended sun-comet vector
  points to where young, small grains of high $\beta$
are expected to be found. In both images, north is up and east is left.
}
\end{figure}
\newline
Fig.~\ref{fig:synsyn} shows the near-nucleus region in more detail and a plot
of synchrones and syndynes as introduced in \cite{finson-probstein1968a} for the same
observation geometry. Syndynes for small $\beta$ and synchrones for old
particles converge towards the direction of the projected comet
orbit. In contrast, recently emitted dust of high $\beta$ is expected to be
found along the direction of the projected sun-comet vector and to the
north-west of it. Given the noisiness of the data, we conclude that we 
cannot detect sufficient evidence for the presence of small, young particles. 
\section{Summary}
\label{sec:summary}
We have observed the dust trail and neckline of comet
67P in April 2004 with the WFI at the ESO/MPG 2.2m
telescope when the comet was at a heliocentric distance of 4.7 AU.
We do not see a coma of small,
recently emitted dust particles around the nucleus. The trail and neckline,
however, are visible over the whole section of the comet orbit covered by the
image ($1.1^\circ$ in mean anomaly). Trail and neckline are separated by
slightly different position angles, in agreement with
theoretical expectation. We have derived lower limits for the
surface brightness of \mbox{$27.0$ mag/arcsec$^2$} close to the nucleus and
\mbox{$28.4$ mag/arcsec$^2$} (im R) further out.
The enhanced surface brightness around the nucleus is
interpreted as the effect of mm/cm-sized dust grains emitted around the
perihelion passage in 2002.  

\section{Acknowledgements}
This work is based on observations made with the MPG/ESO 2.2m telescope at the
La Silla Observatory under programme ID 072.A-9011(A). It has made
use of the USNOFS Image and Catalogue Archive operated by the United States
Naval Observatory, Flagstaff Station.

\bibliographystyle{esa_ref}       
\bibliography{/home/agarwal/Latex/refs}

\begin{thebibliography}{1}

\bibitem{sykes-walker1992a}
{Sykes}, M.~V. and {Walker}, R.~G.
\newblock {Cometary dust trails. I - Survey}.
\newblock {\em Icarus}, 95:180--210, 1992.

\bibitem{sykes-lebofsky1986}
{Sykes}, M.~V., {Lebofsky}, L.~A., {Hunten}, D.~M., and {Low}, F.
\newblock {The discovery of dust trails in the orbits of periodic comets}.
\newblock {\em Science}, 232:1115--1117, 1986.

\bibitem{sykes-hunten1986}
{Sykes}, M.~V., {Hunten}, D.~M., and {Low}, F.~J.
\newblock {Preliminary analysis of cometary dust trails}.
\newblock {\em Advances in Space Research}, 6:67--78, 1986.

\bibitem{kimura-liu1977}
{Kimura}, H. and {Liu}, C.
\newblock {On the structure of cometary dust tails}.
\newblock {\em Chin. Astron.}, 1:235--264, 1977.

\bibitem{erben-schirmer2005}
{Erben}, T., {Schirmer}, M., {Dietrich}, J.~P., {Cordes}, O., {Haberzettl}, L.,
  {Hetterscheidt}, M., {Hildebrandt}, H., {Schmithuesen}, O., {Schneider}, P.,
  {Simon}, P., {Deul}, E., {Hook}, R.~N., {Kaiser}, N., {Radovich}, M.,
  {Benoist}, C., {Nonino}, M., {Olsen}, L.~F., {Prandoni}, I., {Wichmann}, R.,
  {Zaggia}, S., {Bomans}, D., {Dettmar}, R.~J., and {Miralles}, J.~M.
\newblock {GaBoDS: The Garching-Bonn Deep Survey. IV. Methods for the image
  reduction of multi-chip cameras demonstrated on data from the ESO Wide-Field
  Imager}.
\newblock {\em Astronomische Nachrichten}, 326:432--464, 2005.

\bibitem{ishiguro-sarugaku_subm}
{Ishiguro}, M.
\newblock {\em private communication}.

\bibitem{monet-levine2003}
{Monet}, D.~G., {Levine}, S.~E., {Canzian}, B., {Ables}, H.~D., {Bird}, A.~R.,
  {Dahn}, C.~C., {Guetter}, H.~H., {Harris}, H.~C., {Henden}, A.~A., {Leggett},
  S.~K., {Levison}, H.~F., {Luginbuhl}, C.~B., {Martini}, J., {Monet},
  A.~K.~B., {Munn}, J.~A., {Pier}, J.~R., {Rhodes}, A.~R., {Riepe}, B., {Sell},
  S., {Stone}, R.~C., {Vrba}, F.~J., {Walker}, R.~L., {Westerhout}, G.,
  {Brucato}, R.~J., {Reid}, I.~N., {Schoening}, W., {Hartley}, M., {Read},
  M.~A., and {Tritton}, S.~B.
\newblock {The USNO-B Catalog}.
\newblock {\em AJ}, 125:984--993, 2003.

\bibitem{agarwal-mueller2006_inpress}
{Agarwal}, J., {M{\"u}ller}, M., {B{\"o}hnhardt}, H., and {Gr{\"u}n}, E.
\newblock {Modelling the large particle environment of comet
  67P/Churyumov-Gerasimenko}.
\newblock {\em Advances in Space Research}, in press, 2006.

\bibitem{finson-probstein1968a}
{Finson}, M.~L. and {Probstein}, R.~F.
\newblock {A theory of dust comets. I. Model and equations}.
\newblock {\em ApJ}, 154:327--352, 1968.

\end{thebibliography}

\end{document}